 \documentclass[11pt]{article}
 \usepackage{epsfig}
 \usepackage{here}
 \usepackage{amsmath,textcomp}
 \usepackage{bm,graphicx,subfigure} 
    
    \newcommand{\bra}{\langle}
    \newcommand{\ket}{\rangle}
    \newcommand{\bold}[1]{\mathbf{#1}}
    \newcommand{\Half}{\textstyle{\frac{1}{2}}}
    
    \newcommand{\I}{\imath  }
    \newcommand{\D}{\textstyle{\rm d}}
    \newcommand{\E}{\textstyle{\rm e}}
\begin{document}

\begin{center}{\bf\Large  Three-pair final-state interaction 
  in the  
  $\mbox{p}\mbox{p} \boldsymbol{\rightarrow} 
  \mbox{p}\mbox{p}\boldsymbol{\eta}$ reaction close to threshold
  \footnote{$^)$Talk given at COSY-11 meeting, Cracow, 1-3 June, 2004} 
  $\!\!^)$
  }
\end{center}
\vspace{0.5cm}
\begin{center}
   A. Deloff \\[0.1cm]
   {\small \em 
         Institute for Nuclear Studies, Ho\.za 69, 00-681 Warsaw, Poland 
   }
\end{center}
\vspace{0.5cm}
\begin{center}
 \parbox{0.9\textwidth}{
  \small{
    {\bf Abstract:}\
        We present a three-body formalism  describing 
        the final-state interaction effects in the
        $pp\to pp\eta$ reaction close to threshold. 
        We derive a three-body enhancement factor devised 
        in such a way that all three pair-wise interactions
        are regarded on equal footing. The enhancement factor is
        obtained by expanding 
        the three-particle wave function in hyperspherical harmonics.
        It has been shown that close to threshold
        the $p-p$ interaction strongly dominates whereas the $\eta-p$
        interaction gives almost negligible contribution to the calculated
        effective mass spectra.
                Within the presented three-body approach
        it has been possible to reproduce the effective mass distributions  
        at the excitation energy Q=15.5 MeV in good accord with the data.
         }
 }
\end{center}

\vspace{0.5cm}
\section{Introduction}
\begin{figure}
   \begin{center}
   \includegraphics[scale=0.4]{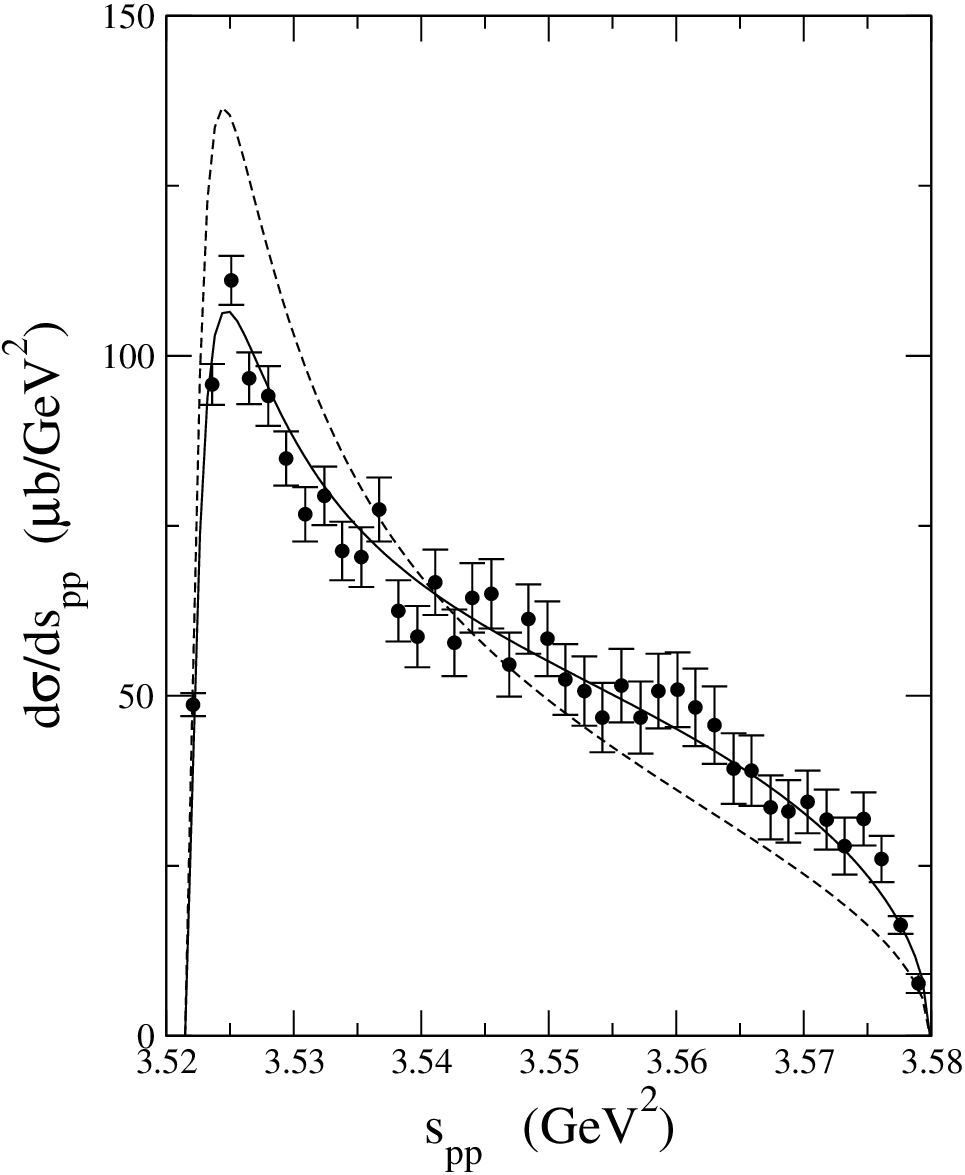}  
   \includegraphics[scale=0.4]{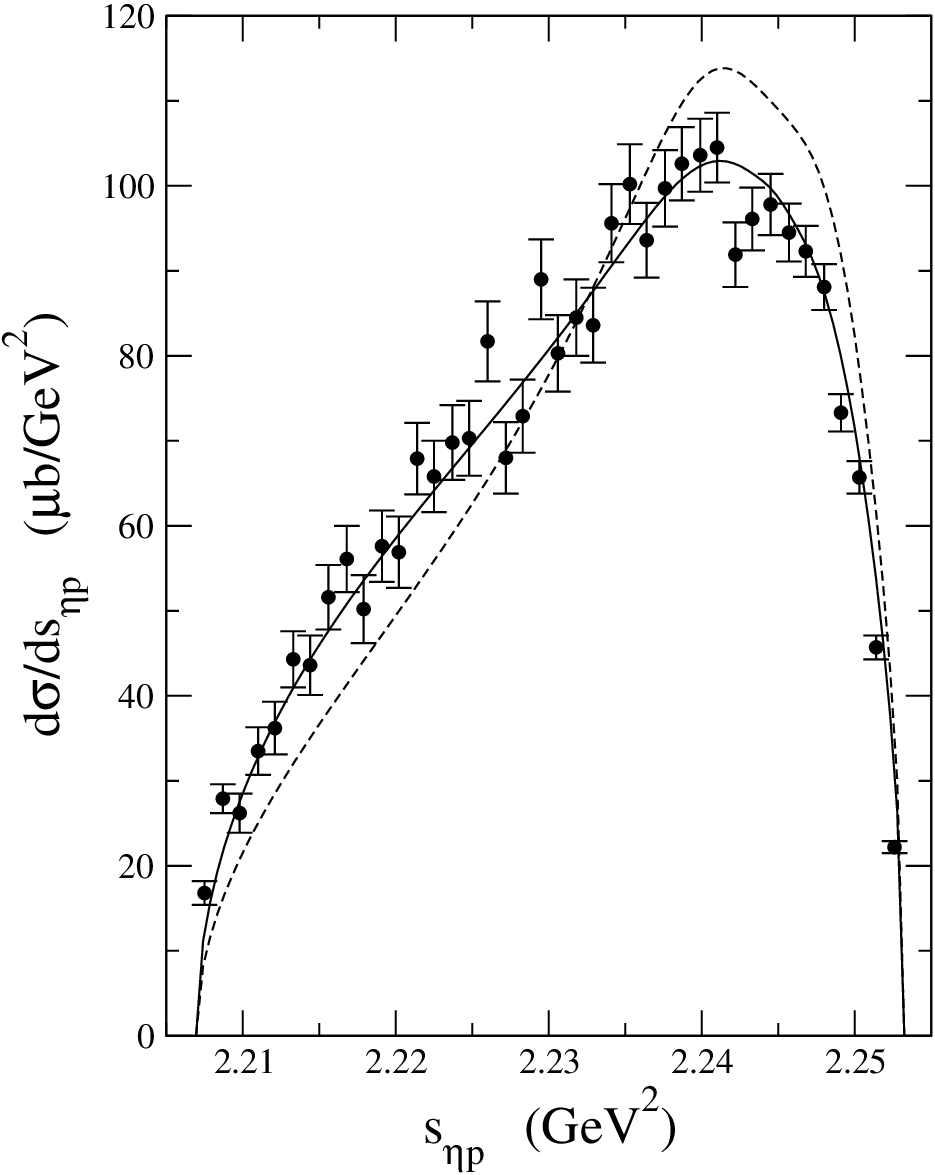}           
 \label{two_body} 
 \caption{ \small
 Invariant mass distributions at Q=15.5 MeV; left panel: (pp)-mass plot; 
 right panel: ($\eta$p)-mass plot.
 The data are from \cite{moskal}
 the calculated curves are from \cite{deloff}.}
 \end{center}
 \end{figure}

 The last decade has seen major advances  in the experimental
 investigation of the near threshold meson production reactions in nucleon-nucleon collisions
 (for a review cf. \cite{review} and \cite{reviewt}).
  In the recent measurements of the $pp\to pp\eta$ reaction a very accurate determination
  of the four-momenta of both 
  outgoing protons allowed for the full reconstruction
  of the kinematics of the final $\eta pp$ state. In consequence, these measurements
  provided in addition to the $\eta$ and the proton angular distributions, also the
  $pp$ and $\eta p$ effective mass distributions \cite{zupran,moskal}. 
  The common feature of the near-threshold
  meson production in proton-proton collisions
  is the dominance of the very strong proton-proton final state interaction
  (FSI).  This effect is 
  clearly visible in the invariant mass distributions: as a prominent peak close 
  to threshold in the (pp)-mass distribution, or as a bump near the end-point
  in the $(\eta p)$-mass spectrum. 
  For sufficiently low excitation energies  the 
  $\mbox{}^3 P_0\to\mbox{}^1 S_0\, s$ transition amplitude 
  becomes necessarily the sole contributor
  to the cross section as
  it is the only amplitude  surviving at threshold. 
  The supposition that this happens at the lowest available
  excitation energy equal Q=15.5 MeV \cite{moskal}
  appears to be quite plausible, especially that
  in a similar experiment \cite{zupran} at Q=15~MeV  
  the measured angular distributions were consistent with isotropy.
  The description in terms of a simple model in which
  a constant $\eta$ production amplitude is multiplied by
  pp FSI enhancement factor, although qualitatively correct, is
  not fully satisfactory in quantitative terms. The
  calculated invariant mass distributions are presented in
  Fig. \ref{two_body} (dashed curves). In order to
  improve the agreement with experiment two possibilities
  have been considered. The most obvious explanation
  admits the contribution from the p-waves, or, more precisely,
   from the $\mbox{}^1\!S_0\to\mbox{}^3\!P_0s$  
   and $\mbox{}^1\!D_2\to\mbox{}^3\!P_2s$ amplitudes. These amplitudes
   have the best chance to show up when the relative momenta
   of the final-state protons take the largest values allowed by the phase space.
   Since at Q=15.5~MeV this sector still overlaps with the peak region of the $\mbox{}^1\!S_0$
   enhancement factor, the s-wave also receives there maximal amplification. Therefore,
   the relative strength of the p-wave amplitudes to be discernible has to be
   quite substantial which in general should be reflected by a pronounced angular
   dependence of the cross section.     
   This difficulty has been thoroughly examined by Nakayama et al. \cite{nakayama} who pointed 
   out that the unwanted angular dependence might still be suppressed under two circumstances:
   (i) if the  $\mbox{}^1\!D_2\to\mbox{}^3\!P_2s$ amplitude was negligibly small 
   so that the angular dependent term was absent,
   or,  (ii) if the lack of angular dependence  resulted from cancellations 
   -- from a destructive interference between
   $\mbox{}^1\!S_0\to\mbox{}^3\!P_0s$  and $\mbox{}^1\!D_2\to\mbox{}^3\!P_2s$ amplitudes.
   Thus, a model basing upon a strong p-wave needs 
   additionally somewhat fortuitous coincidences. According to  
   the second  proposition presented  in \cite{deloff}, the s-wave amplitude dominates
   because it is proportional to the huge FSI enhancement factor whilst the
   p-wave amplitudes are not. Therefore, the latter might be neglected
   explaining  in a natural way the lack of angular dependence. Instead, a weak
   energy dependence is admitted in the production amplitude. Both models are
   capable of improving the agreement with experiment at the expense of introducing
   an adjustable parameter. The best fit from \cite{deloff} is depicted in
   Fig. \ref{two_body} by the full line.
  Since the agreement with experiment presented in \cite{nakayama} is of similar
  quality, polarization experiments are required to discriminate between
  these two models. The prediction of the s-wave model \cite{deloff}
  is that all polarization observables are bound to vanish. Any non-zero
  value for the analyzing power reveals the presence of higher partial
  waves \cite{nakayama}.
  \par
   In the above considerations final-state $\eta$p interaction has been 
  ignored and an interesting question arises how much its inclusion can change
  the resulting invariant mass spectra. The $\eta$p interaction is
  poorly known but there have been suggestions that the corresponding 
  scattering length might be as large as about 1 fm. This is still
  one order of magnitude smaller than the pp scattering length but
  the discrepancies in the invariant mass distribution
  in Fig. \ref{two_body}  are not large either. The purpose of this work
  is to shed some light on the possible role of the $\eta$p
  final-state interaction but to be able to do that we need a
  formalism in which all pair-wise interactions would be regarded on
  equal footing.
  \par
  The plan of our presentation is as follows. 
  In the next Section we start with the well known two-body case 
  recalling the arguments leading to the derivation of the FSI
  enhancement factor. In Section 3 we generalize these ideas
  to the three-body case by utilizing the hypersherical harmonics approach.
  Finally, in Section 4 we verify our model by 
  confronting the obtained results with the experimental data.  
 \section{ Two-body final-state interaction}       
 Since the proton-proton final-state interaction is believed to be the 
 dominant ingredient in the description of 
  the $pp\to pp\eta$ reaction close to threshold, 
 it is logical to begin with the two-body FSI problem.
 The basic idea how to account for final state interaction was 
 put forward 50 years ago by Fermi, Watson, Migdal \cite{Fermi}
 and others (for a review cf.  \cite{Gold})  and is based on the
 observation that in many processes the interaction responsible for carrying the system from 
 the initial to the final state is of such a  short range that
 in the first approximation may be regarded as point like.
 As a prototype one may consider a meson (x) production reaction $NN\to NNx$. To generate
 the meson mass m in nucleon-nucleon collision a large momentum transfer is required
 between the initial and the final nucleons, which is typically of the order
 $\sqrt{Mm}$, with $M$ being the nucleon mass. The corresponding ''range''
 of the production interaction is therefore much shorter than the range of the
 interaction between the two final state nucleons. 
 Although it is perfectly true that 
 the final state NN interaction significantly  distorts the
 NN wave function but in the transition 
 matrix element the contribution from all but the smallest NN separations will be 
 strongly suppressed and the main effect may be attributed to
 the change of  the normalization of the wave function at zero separation. 
 If the non-interacting NN pair is described by a plane wave
 $\E^{-\I\,\bm{k}\cdot\bm{r}}$, where $\bm{k}$ is the relative NN momentum
 ($\hbar=c=1$ units are used hereafter), to account for
 final state interaction the latter must be replaced in the transition matrix element by 
 the complete NN wave function $\Psi^-(\bm{k},\bm{r})^\dagger$ 
 satisfying outgoing spherical wave
 boundary condition at infinity. 
 Nevertheless, for a {\it point-like interaction}, we may set 
 \begin{equation}
 \Psi^-(\bm{k},\bm{r})^\dagger \approx
 \E^{-\I\,\bm{k}\cdot\bm{r}} \; C(k)  
 \label{AA1}
 \end{equation}
 in the matrix element so that the final state interaction will be accounted
 for by multiplying the transition matrix element by the enhancement factor,
 defined as
\begin{equation}
\label{AA2}
 C(k) \equiv  \displaystyle\lim_{r\to 0} \;
 \Psi^+(-\bm{k},\bm{r}) /\E^{-\I\,\bm{k}\cdot\bm{r} }. 
 \end{equation}
The factor $|C(k)|^2$ that appears in the cross section 
represents the ratio of  two probabilities: one of finding the 
interacting NN pair at zero separation,  while the other probability is associated with 
non-interacting particles. By construction, when the final state interaction is turned off,
the enhancement factor will be equal to unity. Expanding both, the numerator and the denominator
on the right hand side of \eqref{AA2} in partial waves, we have
\begin{equation}
 C(k) =  \displaystyle\lim_{r\to 0}\dfrac
 { \sum_{\ell=0}^\infty\; (2\ell+1)\,\I^{-\ell}\,
 \psi_\ell(k,r)/r\;P_\ell(\hat{\bm{k}}\cdot\hat{\bm{r}}) }
 { \sum_{\ell=0}^\infty\; (2\ell+1)\, \I^{-\ell}\;
 j_\ell(kr)\,P_\ell(\hat{\bm{k}}\cdot\hat{\bm{r}}) },
\label{A2a}
\end{equation}
where $\psi_\ell(k,r)\sim r^{\ell+1}$ for small $r$, $j_\ell(kr)$ is spherical Bessel function and $P_\ell(\hat{\bm{k}}\cdot\hat{\bm{r}})$ denotes Legendre polynomial. 
Clearly, in the limit $r\to 0$ in \eqref{A2a}, all higher partial
waves will be suppressed by the centrifugal barrier, and
only the contribution the from s-wave survives. Thus, we obtain a simple formula
\begin{equation}
 C(k) =  \psi_0(k,0)'/k,
\label{C1}
\end{equation}
where prime denotes derivative with respect to $r$, and, 
as apparent from \eqref{C1}, the enhancement factor is 
determined by the slope of the wave function at the origin.
To find this slope we must know the NN s-wave interaction and
for simplicity we shall in the following assume that the 
latter takes the form of a spherically symmetric radial potential.
The shape of this potential may be arbitrary but it must be of a
short range. Given the NN potential,
we can integrate  outward the appropriate wave equation, containing
both the nuclear and the Coulomb potential, 
generating numerically a regular solution $u_0(k,r)$ (i.e. vanishing at the origin)
whose derivative for later convenience is selected to be
\begin{equation}
 u_0(k,0)'=C_0(\eta)\, k,
\label{C2}
\end{equation}
where $\eta$ denotes the Sommerfeld parameter and $C_0(\eta)^2=2\pi\eta/[\exp(2\pi\eta)-1]$ is
the Coulomb barrier penetration factor.
The sought for physical solution $\psi_0(k,r)$ occurring in \eqref{C1} 
which is also regular,
is necessarily proportional to $u_0(k,r)$, and, more explicitly, we have
\begin{equation}
\psi_0(k,r)=[C(k)/C_0(\eta)]\;u_0(k,r). 
\label{C3}
\end{equation}
Now, all we need to calculate  $C(k)$ is the asymptotic expression for
the physical wave function. For $r=R$ with R much bigger than the range of the 
nuclear potential, the physical wave function takes the form  
\begin{equation}
\psi_0(k,R)= F_0(\eta,kR)+f_0(k)\,H^+_0(\eta,kR),
\label{C9}
\end{equation}
where $H^+_0(\eta,kR)= G_0(\eta,kR)+\I\, F_0(\eta,kR)$
with $G_0(\eta,kR)$ and  $F_0(\eta,kR)$ being the standard Coulomb
wave functions defined in \cite{Abram}, 
and $f_0(k)=\sin\delta\,\E^{\I\delta}$ denotes the s-wave scattering 
amplitude with $\delta$ being the s-wave Coulomb distorted phase shift.
 The differentiation of \eqref{C9} with respect to R, provides us with
 a second condition for the derivatives but
 it should be noted that  $u_0(k,R)$ and $u(k,R)'$ 
 occurring in these two matching conditions
are to be regarded as known quantities. Indeed,  they are fully specified by 
the boundary condition at the origin \eqref{C2} and can be 
either calculated analytically, or obtained by numerical methods.  
Therefore,  we end up with  two algebraic equations in which the two unknowns
are the enhancement factor
$C(k)$ and the scattering amplitude $f_0(k)$ and the respective solutions,   
can be conveniently written as
\begin{equation}
C(k)=\dfrac{k\,C_0(\eta)}{w[H^+_0(\eta,kR),u_0(k,R)]},
\label{C10}
\end{equation}
and
\begin{equation}
f_0(k)=-\dfrac{w[  F_0(\eta,kR),u_0(k,R)]}
              {w[H^+_0(\eta,kR),u_0(k,R)]},
 \label{C10a}
 \end{equation}
 where the symbol $w[f,g]$ denotes the Wronskian 
 defined as $w[f,g]\equiv fg'-f'g$. 
 The specific Wronskian occurring
 in the denominators of  \eqref{C10}--\eqref{C10a}  
 has been referred to as the Jost function  \cite{Gold}.
 Thus, given the potential, the two-body enhancement factor
 can be obtained from \eqref{C10}.
 \section{ Three-body final-state interaction}
 We wish now to extend the ideas outlined above to the three-body case.
 We assume that the pair-wise short range interaction between the particles 
 is strong but the system is Borromean, i.e. {\em  
 no binary bound state may exist in the three-body system under consideration}.
 A formal theory of the continuum in a Borromean system was developed in \cite{HH} 
 using the same hyperspherical harmonics method which has been widely employed for the 
 investigation of bound states and specifically the halo nuclei \cite{HHH}. In this paper
 we follow this approach to derive the enhancement factor for three interacting particles.
 The wave functions in the continuum are solutions to the three-body problem satisfying
 the correct boundary conditions at infinity
 where the three-body asymptotics is  most naturally
 expressed in terms of the rotationally and permutationally invariant hyperradius $\rho$
 defined as the square root of the sum of squares of the inter-particle distances.
 Since the hyperradius reflects the size of the three-body system, 
 similarly as in the two-body case, the enhancement factor may be obtained
 as the limiting value $\rho\to 0$  of the three-body wave function. 
 The method of hyperspherical harmonics has been well documented in the literature, 
 but to make this paper self-contained we wish to summarize briefly the theoretical
 framework necessary to treat the continuum of a Borromean system.
 \subsection{Coordinate sets and hyperspherical harmonics}
  For assigned particle positions 
  $(\bold{r}_1, \bold{r}_2,\bold{r}_3)$ and masses $(m_1, m_2, m_3)$,
 the translationally invariant normalized sets of Jacobi coordinates
 $\bold{x}_i,\bold{y}_i$ are defined, as follows
 \begin{subequations}
 \begin{eqnarray}
     \bold{x}_i &=& \sqrt{\dfrac{m_j\, m_k}{(m_j + m_k)\,\mu}}\,
     (\bold{r}_j-\bold{r}_k), 
   \label{A1:a}
     \\
     \bold{y}_i &=& \sqrt{ \dfrac{m_i\,(m_j+m_k)}{\mu\,M} } 
     \,\left (\bold{r}_i -
       \dfrac{m_j\,\bold{r}_j+m_k\,\bold{r}_k}{m_j+m_k} \right ),   
   \label{A1:b}
   \\
     \bold{R} &=&  ( m_i \bold{r}_i + m_j \bold{r}_j +m_k \bold{r}_k )/M,
   \label{A1:c}
 \end{eqnarray}
   \label{A1}
 \end{subequations}
  where $\{i,j,k\}$ is a permutation of the particle labels $\{1,2,3\}$,
  $M=m_1+m_2+m_3$, and
  $ \mu $ denotes an arbitrary mass which just sets the mass scale. 
  Each of the three equivalent pairs  $(\bold{x}_i,\bold{y}_i)$ 
  together with the center of mass coordinate $\bold{R}$
  describes the system. The transformation between different sets of Jacobi
  coordinates is referred to as a kinematical rotation and takes the form 
   \begin{equation}
    \binom{\bold{x}_j}{\bold{y}_j}=
           \begin{pmatrix}
                   \text{-}\cos\omega_{ij}& \text{ }\sin\omega_{ij}\\      
                   \text{-}\sin\omega_{ij}& \text{-}\cos\omega_{ij}        
           \end{pmatrix}
    \binom{\bold{x}_i}{\bold{y}_i}
           \label{A1p}
   \end{equation}
  where the rotation angle, confined by $-\pi/2 \le \omega_{ij} \le \pi/2$,
  is given by
  \begin{equation}
  \omega_{ij}=\arctan \left[ \sigma \{i,j,k\}
  \sqrt{M\, m_k /m_i\, m_j } \right ],
  \label{A2p}
  \end{equation}
  with $\sigma \{i,j,k\}$ denoting the sign of the permutation $\{i,j,k\}$.%
  \par
  The Jacobi momenta $\bold{k}_i, \bold{q}_i$ and
  $\bold{P}$, canonically conjugate to $\bold{x}_i, \bold{y}_i$ and $\bold{R}$,
  respectively, are defined by the relations,
  \begin{subequations}
  \begin{eqnarray}
  \bold{k}_i &=& \sqrt{\dfrac{m_j\, m_k\,\mu}{m_j+m_k}}\,
  \left(\frac{\bold{p}_j}{m_j}-\frac{\bold{p}_k}{m_k}\right),
  \label{A3p:a} \\
  \bold{q}_i &=& \sqrt{\dfrac{(m_{j}+m_{k})m_i\mu}
   { M } }\,    
  \left(\frac{\bold{p}_i}{m_i}-
  \frac{\bold{p}_j+\bold{p}_{k}}{m_j+m_k}\right),
  \label{A3p:b} \\
  \bold{P} &=& \bold{p}_1 + \bold{p}_2 + \bold{p}_3,
  \label{A3p:c}
  \end{eqnarray}
  \label{A3p}
  \end{subequations}
  where $\bold{p}_{i}\; (i=1,2,3)$ are the laboratory frame momenta.%
  \par
  Instead of the Jacobi coordinates we shall use hyperspherical coordinates
  comprising of a hyperradius $\rho$ and five angles. 
  The hyperradius $\rho$ determines the size of a three--body system and
  is invariant with respect to kinematic rotations
 \begin{equation}
 \rho^2=x_{i}^2+y_{i}^2, \qquad i=1,2,3.
 \label{A2}
 \end{equation}
 The five angular variables forming a five-dimensional solid angle
 $\Omega_\rho$ include the usual angles $(\theta_x^i,\phi_x^i)$,
 $(\theta_y^i,\phi_y^i)$ specifying the unit vectors $\hat{{\bf x}}_i$,
 $\hat{{\bf y}}_i$ and these are supplemented by the
 hyperangle $\alpha_i$ defined by the equalities
 \begin{equation}
   x_i = \rho\sin\alpha_i,\qquad
   y_i = \rho\cos\alpha_i,
  \label{A3}
 \end{equation}
 where $0 \le \alpha_i \le \pi/2$. The 
 five-dimensional volume element d$\Omega_{\rho}$  is
 \begin{equation}
   \D \Omega_{\rho}=
   \sin^2\alpha_i\,\cos^2\alpha_i\, \D \alpha_i\, \D \hat{\bf x}_i\,\D \hat{\bf y}_i.
 \label{A4}
 \end{equation}
 For the conjugate momenta, 
 we proceed in a similar fashion introducing
 the hypermomentum $\kappa$ and
 the associated hyperangle $\beta_i$
 \begin{equation}
   k_i = \kappa\sin\beta_i,\qquad
   q_i = \kappa\cos\beta_i,
 \label{A4p}
 \end{equation}
 where owing to the energy conservation $Q=\kappa^2/2\mu$. 
 \par
 In the six-dimensional space, the kinetic energy operator $\hat{T}$ takes 
 a separable form
 \begin{equation}
   \hat{T}=
 -\frac{1}{2\mu} \left (\dfrac{\partial^{2}}{\partial\rho^{2}}+
 \dfrac{5}{\rho} \dfrac{\partial}{\partial\rho}-
 \dfrac{1}{\rho^{2}} \hat{K}^{2} \right ),
 \label{A5p}
 \end{equation}
 where the hypermomentum operator $\hat{K}$ is the generator of rotations
 in the six-dimensional space. The operator $\hat{K}^{2}$
 takes the form
 \begin{equation}
 \hat{K}^{2}=-\dfrac{ \partial^{2} }{ \partial\alpha_i^2 }
 -4\cot (2\alpha_i) \dfrac{ \partial }{ \partial\alpha_i}
 +\dfrac{ \hat{\bm{l}}_i^2  }{ \sin^{2} \alpha_i }
 +\dfrac{\hat{ \bm{\lambda}}_i^2  }{ \cos^{2} \alpha_i },
 \label{A6p}
 \end{equation}
 where the angular momentum operators $\hat{\bm{l}}_i$ and $\hat{\bm{\lambda}}_i$
 occurring in \eqref{A6p}, are defined as
 \begin{equation}
         \hat{ \bm{l}}_i=-\I\,\bm{x}_i\times\,\dfrac{\partial}{\partial\,\bm{x}_i},\qquad
         \hat{ \bm{\lambda}}_i=
         -\I\,\bm{y}_i\times\,\dfrac{\partial}{\partial\,\bm{y}_i},
         \label{A6pp}
 \end{equation}
 and $\hat{\bm{l}}_i^2$ and $\hat{\bm{\lambda}}_i^2$ have eigenvalues $\ell(\ell+1)$ and
 $\lambda(\lambda+1)$, respectively. The operator \eqref{A6p}
 has eigenvalues $K(K+4)$ where $K=2n+\ell + \lambda$ for integer $n$.
 The quantum number $K$ has been dubbed hypermomentum, its value
 is the same in all three Jacobi systems, and 
 the corresponding eigenfunctions are known as hyperspherical harmonics
 (abbreviated HH hereafter).%
 \par
 From now on we choose  set 3 as the basic one
 and to simplify  notation in the following we will suppress the label referring
 to our particular choice of the Jacobi system. 
 The HH have the explicit form
 \begin{equation}
 {\cal Y}_{KLM_L}^{\ell\lambda} \left( \Omega_{\rho} \right) = \chi_{K}^{\ell\lambda}
 ( \alpha ) 
 \left[ Y_{\ell} (\hat{\bf x}) \otimes Y_{\lambda} (\hat{\bf y})
 \right]_{LM_L},
 \label{A6}
 \end{equation}
 where $L$ is the total angular momentum resulting from vector addition
 $\bm{L}=\bm{l}+\bm{\lambda}$. In \eqref{A6} the symbol
  $\Omega_{\rho}$ denotes collectively all five angular variables with 
 $Y_{lm}$ being the usual spherical harmonics.  
 The square bracket in \eqref{A6} indicates
 vector coupling of $\bold{l}$ and $\bm{\lambda}$ 
 producing a total angular momentum $\bold{L}$ and  the appropriate    
 quantum numbers associated with the latter operator are $L$ and $M_L$.
 In \eqref{A6}, the hyperangular eigenfunctions are 
 \begin{equation*} 
 \chi_{K}^{\ell\lambda} ( \alpha) = N_K^{\ell\lambda}
 \;\sin^\ell\alpha \,\cos^\lambda\alpha \,
 P_n^{\ell+\!\!\Half,\scriptstyle\lambda+\!\!\Half} (\cos 2\alpha),
 \end{equation*}
 where $P_n^{\alpha,\beta} (x)$ are the Jacobi polynomials (cf.  \cite{Abram})
  and $N_K^{\ell\lambda}$ is the normalization constant.
 The HH in \eqref{A6} are
 orthonormalized using the volume element~\eqref{A4} and
 $\alpha$ stands for any of the $\alpha_i$. 
 \par
 For clarity reasons,  to avoid unnecessary complications,
 we are going  to restrict our considerations to the
 simplest case when the particles have no internal degrees of freedom.
 When the  particles are moving freely the spacial part of the
 three-particle wave-function will be described in the c.m. frame
 by a plane wave, whose HH expansion can be written as
 \begin{equation}
         \E^{ \I \bold{k}\cdot\bold{x} +\I\bold{q}\cdot\bold{y} }
 =\dfrac{(2\pi)^{3}} {(\kappa\rho)^{5/2}} \,\displaystyle
 \sum_{L M_L K\ell\lambda}
 \; {\cal Y}^{\ell\lambda}_{KLM_L}(\Omega_\rho)
 \;\I^K\, \sqrt{\kappa\rho}\,J_{K+2}(\kappa\rho)
 \; {\cal Y}^{\ell\lambda}_{KLM_L}(\Omega_\kappa)^{\dagger},
 \label{A7}
 \end{equation}
 where $J_{K+2}$ denotes Bessel function of integer order and
 solid angle $\Omega_{\kappa}$ comprises five angular variables that specify 
 the directions of the incident momenta  in the six-dimensional  space.
 It is worth noting that since the radial part in \eqref{A7}
 depends solely upon $K$, the plane wave is an invariant under
 six-dimensional rotations. The two-body interactions break up
 this invariance and for interacting
 particles the three-particle wave function
 in the continuum $\Psi_{\bold{k},\bold{q}}(\bold{x},\bold{y})$
 will have a more complicated
 HH expansion. For the simplest case of pair-wise central
 potentials depending upon the particle separations, we have
 \begin{equation}
         \Psi_{\bold{k},\bold{q}}(\bold{x},\bold{y})=
         \dfrac{(2\pi)^3}{(\kappa\rho)^{5/2}}
         \sum
 {\cal Y}^{\ell \lambda}_{K L M_L}(\Omega_\rho)
 \;\psi^L_{K\ell\lambda,\tilde{K}\tilde{\ell}\tilde{\lambda}} (\kappa,\rho)
         \; {\cal Y}^{\tilde{\ell} \tilde{\lambda}}_{\tilde{K}L M_L}(\Omega_\kappa)^\dagger
 \label{A8}
 \end{equation}
 where the summation indices are
    $L,M_L, K,\ell,\lambda,\tilde{K},\tilde{\ell},\tilde{\lambda}$
 and the normalization condition reads
 \begin{equation}
         \int
         \Psi_{\bold{k'},\bold{q'}}(\bold{x},\bold{y})^\dagger\;
         \Psi_{\bold{k},\bold{q}}(\bold{x},\bold{y})\;\D^3x\,\D^3y =
         \delta(\bold{k}'-\bold{k})\,\delta(\bold{q}'-\bold{q}).
         \label{A8'}
 \end{equation}
 In the expansion \eqref{A8}, only the total
 three-particle angular momentum L is a good quantum number
 and the as yet unspecified hyperradial part 
 $\psi^L_{K\ell\lambda,\tilde{K}\tilde{\ell}\tilde{\lambda}} (\kappa,\rho)$
 must be determined from the underlying three-body dynamics.
 The dependence upon the tilded indices, associated with the incident momenta, 
 enters  solely {\it via} the asymptotic boundary condition at $\rho \to \infty$.
 The full wave-function
 $\Psi_{\bold{k},\bold{q}}(\bold{x},\bold{y})$
 is a solution to the three-body Schr\"odinger equation
 \begin{eqnarray}
 (\hat{T} + \hat{V} - Q) 
 \Psi_{\bold{k},\bold{q}}(\bold{x},\bold{y})=0,
 \label{A9}
 \end{eqnarray}
 with
 $\hat{V} = \hat{V}_{12} + \hat{V}_{23} + \hat{V}_{31}$ where
 $\hat{V}_{ij}$
 stands for the pair-wise interaction between particles
 $i$ and $j$. Inserting \eqref{A8} in \eqref{A9} and projecting onto the
 hyperangular part of the wave function, results in an infinite set 
 of coupled systems of differential equations 
 enumerated by  the conserved total angular momentum $L$ -- the only quantum
 number that does not mix. For an assigned $L$, we have
 \begin{equation}
 \begin{split}
                 \left\{\dfrac{\D^{2}}{\D\rho^{2}} -\dfrac{ (K+3/2)(K+5/2)}{\rho^2}
 +\kappa^{2} \right\}
 \;\psi^L_{K\ell\lambda,\tilde{K}\tilde{\ell}\tilde{\lambda}} (\kappa,\rho)
  =\hspace*{0.2\textwidth} \\
  =2\mu\sum_{K'\ell'\lambda'}
 \bra K'\ell' \lambda' |\hat{V}|K \ell \lambda  \ket \;
 \;\psi^L_{K'\ell'\lambda',\tilde{K}\tilde{\ell}\tilde{\lambda}} (\kappa,\rho),
 \end{split}
 \label{A10}
 \end{equation}
 and we are left with a multi-channel situation where each channel is
 specified by three quantum numbers $\{K\ell\lambda\}$. 
 Each system of equations \eqref{A10} with $L=0,1,2,\dots$ is infinite
 because there is no upper limit for $K$, and, therefore,
 for practical reasons $K$ must be truncated at some finite value $K_{max}$ so that
 the orbital momenta $\ell$ and $\lambda$ are thereby restricted to
 vary in finite limits. The value of $K_{max}$ determines the order of the
 approximation and must be be large enough to ensure the convergence of the method.
 It should be noted that
 in \eqref{A10} the potential term has been sandwiched between the HH functions 
 ${\cal Y}^{\ell \lambda}_{KLM_{L}}(\Omega_\rho)$
 and after integration over the five angular variables $\Omega_\rho$,
 these matrix elements  depend solely upon the hyperadius $\rho$
 \begin{equation}
 \bra K'\ell' \lambda' |\hat{V}|K \ell \lambda  \ket 
 =\int
 {\cal Y}^{\ell' \lambda'}_{K' L M_L}(\Omega_\rho)^\dagger
 \;\sum_{j=1}^3 \sum_{i \neq j}^3 V_{ij}(\rho,\Omega_\rho)
         \; {\cal Y}^{\ell\lambda}_{KL M_L}(\Omega_\rho)\;\D\,\Omega_\rho.
         \label{A10a}
 \end{equation}
 With the adopted here set 3 of HH,
 the computation of the potential matrix element $\hat{V}_{12}$ 
 is relatively straightforward because the separation vector 
 between particles 1 and 2 is proportional to
 $|\bold{x}|$ and the calculation of the potential
 matrix involves a single integration. By contrast,
 in the two remaining potentials
 the corresponding separations, $r_{23}$ and $r_{31}$, respectively, 
 are linear combinations of the two vectors $\bold{x}$ and $\bold{y}$
 which unavoidably leads to five-dimensional integrations in
 the calculation of the potential matrix.  Fortunately, there is
 a very efficient procedure to overcome this difficulty, based on the 
 observation that all HH with fixed $(K,L)$ values
 defined with respect to  set $i$ (cf. \eqref{A1})
 are linear combinations of HH belonging
 a set $j$ and this transformation is effected by means of the so called
 Raynal-Revai (RR) coefficients \cite{RR}, {\it viz.}
 \begin{equation}
 {\cal Y}^{\ell_i \lambda_i}_{KLM_{L}}
 (\alpha_i,\bold{\hat{x}}_i,\bold{\hat{y}}_i)=
 \sum_{ \ell_j\,\lambda_j }
 \bra \ell_j\,\lambda_j| \ell_i\, \lambda_i \ket_{KL}\;
 {\cal Y}^{\ell_j \lambda_j}_{KLM_{L}}
 (\alpha_j,\bold{\hat{x}}_j,\bold{\hat{y}}_j),
 \label{A11}
 \end{equation}
 where the RR coefficients $\bra \ell_j\,\lambda_j| \ell_i\, \lambda_i \ket_{KL}$
 are functions of the angle specifying
 the kinematic rotation $i\to j$ given in \eqref{A2p}. 
 With the aid of \eqref{A11}, the potential matrix of $\hat{V}_{23}$
 may be computed in basis $1$ where this task is simple
 and subsequently transformed to basis $3$.%
 \par
 The radial wave functions must be regular at $\rho=0$ and the boundary
 condition imposed on the solutions of \eqref{A10} is    
 \begin{equation}
         \psi_{ K\ell\lambda,\tilde{K}\tilde{\ell}\tilde{\lambda} }^L (\kappa,\rho)
         \sim \rho^{K+5/2}\qquad \text{for}\qquad \rho \to 0.
         \label{A11bis}
 \end{equation}
 For $\rho\to\infty$ the potential term in \eqref{A10} goes to zero
 so that in this limit the system of equations \eqref{A10} becomes  decoupled. 
 In absence of the potential term, the asymptotic solutions
 of the radial equations are well known, they are given in terms of the Bessel and Neuman
 functions  \cite{Abram} $J_{K+2}(\kappa\rho)$ and
 $Y_{K+2}(\kappa\rho)$, respectively. 
 With the Coulomb interaction present,
 for large $\rho$ strong interaction potentials become negligible,
 and  we set
 \begin{equation}
         \lim_{\rho\to\infty}
         \psi_{ K\ell\lambda,\tilde{K}\tilde{\ell}\tilde{\lambda} }^L (\kappa,\rho)
         =\delta_{K\tilde{K}}\,\delta_{\lambda\tilde{\lambda}}\,\delta_{\ell\tilde{\ell}}
                 \;F_{\cal L}(\eta,\kappa\rho)
         +\bra K\ell\lambda| {\cal T}^L| \tilde{K}\tilde{\ell}\tilde{\lambda} \ket 
         \;H^+_{\cal L}(\eta, \kappa\rho),
 \label{A12}
 \end{equation}
 with
 \begin{equation}
         H^+_{\cal L}(\eta,\kappa\rho)= G_{\cal L}(\eta,\kappa\rho)
                 +\I\,F_{\cal L}(\eta, \kappa\rho).
         \label{A12bis}
 \end{equation}
 where we have extended the standard definition of Coulomb wave functions
 so that the orbital quantum number $\ell$ which takes on integer values 
 is replaced by ${\cal L}=K+3/2$ with half-integer ${\cal L}$.
 The T-matrix occurring in \eqref{A12} 
        $ \bra K\ell\lambda| {\cal T}^L| \tilde{K}\tilde{\ell}\tilde{\lambda} \ket $
 describes  $3\to 3$ scattering processes and is similar to that encountered in the two-body
 multichannel problem. In particular, it 
 can be diagonalized which allows to determine the appropriate eigenphases $\delta_{LK\ell\lambda}$
 whose rapid variation with energy indicates a resonance.
 \subsection{Three-body final-state interaction enhancement factor}
 The formalism developed in the preceding subsection will be now applied
 to calculate the enhancement factor describing final-state interaction between
 three particles in a Borromean system. The approximation scheme will be
  based on the assumption that all interactions are of a short
 range and therefore the most important effect comes from the distortion
 of the wave function at very small separations. Bigger separation
 region is strongly suppressed by the small size of the interaction volume
 and gives relatively small contribution to the overlap integrals
 representing reaction transition amplitudes. Thus, similarly as in
 the two-body case, we shall define the enhancement factor as the
 limit reached by the square of the absolute value of the ratio
 -- the full three-body wave function divided by the plane wave -- 
 when the size of the system  represented
 by $\rho$ goes to zero. It is apparent from \eqref{A10} that 
 the role of the centrifugal barrier plays
 now the quantum number $K$ and the leading term
 in the wave function in the limit $\rho\to 0$  has
 $K=0$  implying that  $\ell = \lambda = 0$  and  $L=0$.  
 Therefore, we confine our attention solely to the $L=0$
 set \eqref{A10} where
 significant simplifications take place,  namely we have $\lambda=\ell$ and
 $K$  necessarily   even: $K=2(n+\ell)$.
 \par
 Formally, the three-particle enhancement factor will be defined as the limit
 \begin{equation}
         F(\bm{k},\bm{q})=\lim_{\rho\to 0}\left|\dfrac
         {\Psi_{\bold{k},\bold{q}}(\bold{x},\bold{y})}
         { \E^{ \I \bold{k}\cdot\bold{x} +\I\bold{q}\cdot\bold{y} } }\right|^2,
         \label{B1}
 \end{equation}
 which by making  use of \eqref{A7} and \eqref{A8}, simplifies to the form
 \begin{equation}
         F(\bm{k},\bm{q})=
         \lim_{\rho\to 0}
         \left| \sum_{ \tilde{K}, \tilde{\ell} }
 \;\dfrac{
 \psi^0_{000,\tilde{K}\tilde{\ell}\tilde{\ell}} (\kappa,\rho)}
 {\sqrt{\kappa\rho}\,J_2(\kappa\rho)}\;
         \; {\cal Y}^{\tilde{\ell} \tilde{\ell}}_{\tilde{K}00}(\Omega_\kappa)^\dagger
         \right|^2,
         \label{B2}
 \end{equation}
 where we have dropped the uninteresting constant factor  $ |{\cal Y}^{00}_{000}(\Omega_\rho)|^2$.
 To obtain the enhancement factor it is sufficient to calculate the radial wave function  
 $\psi^0_{000,\tilde{K}\tilde{\ell}\tilde{\ell}} (\kappa,\rho)$ for small $\rho$
 and take the limit indicated in \eqref{B2}.%
 \par
 The calculation of the radial wave function is carried out similarly as for the
 two-body case except that scalar quantities now need to be  replaced by matrices
 in channel space. These channels are specified by two quantum numbers ${K\ell}$
 but to simplify the notation it is possible to use a single integer
 $\nu=\Half\;(\Half K)(\Half K+1)+\ell+1$ enumerating all the states under consideration
 ($\nu$=1,2,3\ldots).
 Truncating the infinite sequence of $K$ values at some assigned $K_{max}$,
 the total number of channels $N$ is
 \begin{equation}
                N= \Half (\Half K_{max}+1)(\Half K_{max}+2),
         \label{B4}
 \end{equation}
 and we have to solve $N$  second order differential equations given in \eqref{A10}.
 The physical solution of \eqref{A10}
 can be envisaged as a column vector in channel space. To obtain the physical solution 
 we must first generate numerically  $N$ linearly independent particular solutions vanishing at the origin
 with an arbitrary slope at $\rho=0$. These $N$ solutions can be grouped together
 to form the $N$ columns of a single $N\times N$ square matrix $\bm{\Phi}$
 (hereafter we denote matrices by boldface symbols).
 In practice, this matrix may be obtained by solving $N$ times the system of 
 equations \eqref{A10} 
 imposing for small $\rho$  the boundary condition   
 \begin{equation}
         \Phi_{K\ell,\tilde{K}\tilde{\ell}}(\kappa,\rho)= \delta_{K\tilde{K}}\,\delta_{\ell\tilde{\ell}}
         \,(\kappa \rho)^{K+5/2},
         \label{B5}
 \end{equation}
 so that these particular solutions will be enumerated by two quantum numbers $\tilde{K}\tilde{\ell}$.
 Since the columns of $\bm{\Phi}$ are linearly independent, they span the space of all possible
 slopes. Therefore, any column vector solution $\psi$ of \eqref{A10} must be expressible
 as some linear combination of the N columns of $\bm{\Phi}$, and we have
 \begin{equation}
         \psi(\rho)=\bm{\Phi}(\rho)\cdot c,
         \label{B6}
 \end{equation}
 with $c$ denoting a column vector of constant coefficients.%
 \par
 The matrix of physical solutions occurring in \eqref{B2} differs from  $\bm{\Phi}$ at
 most by a constant matrix and the latter will be explicitly determined by making use
 of the boundary condition for asymptotic $\rho$.
 For large $\rho$ the physical solution must be of the form
 \begin{equation}
         \bm{\Psi}(\kappa,\rho) = \bm{F}(\kappa\rho)+\bm{H}^+(\kappa\rho)\cdot\bm{T}(\kappa),
         \label{B6a}
 \end{equation}
 where $\bm{T}$ is the true $3\to 3$ scattering matrix. In formula \eqref{B6a}  
 $\bm{F}$ denotes a diagonal matrix containing the regular solutions 
 $F_{\cal L}(\eta,\kappa\rho)$
 of \eqref{A10} in absence of the strong interaction, 
 whereas $\bm{H}^+$ is another diagonal matrix
 containing outgoing hyperspherical waves 
   $G_{\cal L}(\eta, \kappa\rho)+ \I\,F_{\cal L}(\eta,\kappa\rho)$.
 Since $\bm{\Phi}$ is proportional to  $\bm{\Psi}$, for some large $\rho=\rho_m$, we may set
 the following matching condition for the wave functions and their derivatives
 \begin{subequations}
         \begin{eqnarray}
   \bm{\Phi}(\kappa,\rho_m)\cdot \bm{C}(\kappa) & =& \bm{F}(\kappa\rho_m)+\bm{H}^+(\kappa\rho_m)\cdot\bm{T}(\kappa),
                 \label{B7:a}\\
   \bm{\Phi}(\kappa,\rho_m)'\cdot\bm{C}(\kappa) & =& \bm{F}(\kappa\rho_m)'+\bm{H}^+(\kappa\rho_m)'\cdot\bm{T}(\kappa),
                 \label{B7:b}
         \end{eqnarray}
                 \label{B7}
 \end{subequations}
 where prime denotes derivative with respect to $\rho$ and
 $\bm{C}(\kappa)$ is a constant (albeit energy dependent) matrix to be determined.
 Eliminating $\bm{T}$ between the  two matrix equations \eqref{B7}, we obtain
 \begin{equation}
         \bm{C}(\kappa)=\dfrac{2\kappa}{\pi}[
         \bm{H}^+(\kappa\rho_m)\cdot\bm{\Phi}(\kappa,\rho_m)'-
         \bm{H}^+(\kappa\rho_m)'\cdot\bm{\Phi}(\kappa,\rho_m)]^{-1},
         \label{B8}
 \end{equation}
 and $\bm{C}(\kappa)$ is seen to be proportional to the inverse of the Jost matrix.  
 Clearly,  formula \eqref{B8} may be viewed as a three-body extension of \eqref{C10}. 
 Because  the matrix 
 $\bm{C}(\kappa)$ involves the inverse of the Jost determinant it is bound 
 to have the same singularity structure as the T-matrix.%
 \par
 The physical solution can be
 now expressed entirely in terms of $\bm{\Phi}$
 \begin{equation}
         \bm{\Psi}(\kappa,\rho)=\bm{\Phi}(\kappa,\rho)\cdot\bm{C}(\kappa)
         \label{B9}
 \end{equation}
 and the limiting procedure $\rho\to 0$ in \eqref{B2} may be effected with the aid of \eqref{B5}. 
 Similarly as in the two-body case, the only terms in the summations
 giving non-vanishing contribution are those with $K=0$. 
 Leaving out irrelevant numerical factor, the enhancement factor in \eqref{B2} takes the form
 \begin{equation}
         F(\bm{k},\bm{q})=
         F(-\bm{k},-\bm{q})=\left|\sum_{K\ell}c_{K\ell}(\kappa)
         \; {\cal Y}^{\ell\ell}_{K00}(\Omega_\kappa)
         \right|^2,
         \label{B10}
 \end{equation}
 where the expansion coefficients 
 $c_{K\ell}(\kappa)$ are provided as the first row of the matrix $\bm{C}(\kappa)$ given in 
 \eqref{B8}.%
 \par
 Formula \eqref{B10} gives the enhancement factor in the form of an
 expansion in an orthonormal set of momentum dependent HH functions 
 which turns out to be quite useful in effecting phase space
 integrations. We shall illustrate this point for the case when the transition matrix is
 assumed to be constant so that the cross section is essentially determined by the enhancement factor.
 The integration over the five angles $\D\Omega_\kappa$, yields the total cross section 
 \begin{equation}
         \sigma(\kappa)\sim \dfrac{\kappa^4}{f(\kappa)}\;\sum_{K\ell}|c_{K\ell}(\kappa)|^2,
         \label{B11}
 \end{equation}
 where $\kappa^4$ results from phase space 
 and $f(\kappa)$ denotes the incident flux factor. The summation over $K$ extends
 over all even numbers from 0 up to $K_{max}$ and in the following we take it
 to be the value at which the convergence has been attained.%
 \par
 When the incident energy is fixed (i.e, $\kappa$ is a constant),
 the quantities of interest are usually
 the effective mass distributions, or equivalently, the corresponding
 kinetic energy distributions of the
 different pairs in their c.m frame. 
 Such distributions are obtained  here by integrating only over the directions
 of $\bm{k}$ and $\bm{q}$  retaining the dependence upon the two remaining variables,
 $\kappa$ and $\beta$. 
 With our choice of 3 as the basic Jacobi set, the distribution of the center of mass energy
 $T_3$ of the pair $\{12\}$, takes particularly simple form
 \begin{equation}
         \dfrac{\D\sigma}{\D T_3}=\sqrt{T_3(Q-T_3)}\,
         \sum_{K'K\ell}c_{K'\ell}(\kappa)^*\,c_{K\ell}(\kappa)\,
         \chi^{\ell\ell}_{K'}(\beta)\,\chi^{\ell\ell}_K(\beta),
         \label{B12}
 \end{equation}
 where $T_3=Q\,\sin^2 \beta$ and the square root factor is a remanet of the phase space.
 All other factors that depend only upon $\kappa$ have been dropped as they
 may be absorbed in the arbitrary normalization constant. The calculation of
 c.m kinetic energy distributions of the two remaining pairs, i.e  $\{23\}$ and $\{31\}$, respectively,
 may be carried out in exactly the same manner but the HH occurring in \eqref{B10}
 need to be first transformed to the appropriate Jacobi system
 by applying a kinematical rotation.
 Thus, using the transformation \eqref{A11} in \eqref{B10},
 the distribution of the kinetic energy of the $\{23\}$ pair, $T_1$, is
 \begin{equation}
         \begin{split}
                 \dfrac{\D\sigma}{\D T_1}=\sqrt{T_1(Q-T_1)}\;
         \sum_{K'K\ell^\prime_3\ell_3\ell_1}c_{K'\ell^\prime_3}(\kappa)^*\,c_{K\ell_3}(\kappa)\,
         \chi^{\ell_1\ell_1}_{K'}(\beta_1)\,\chi^{\ell_1\ell_1}_K(\beta_1)
                \times \\ \times 
       \;\bra\ell_1\ell_1|\ell_3\ell_3\ket_{K0}\;
         \bra\ell_3^\prime\ell_3^\prime|\ell_1\ell_1\ket_{K'0},
                 \hspace*{0.1\textwidth}
 \end{split}
         \label{B13}
 \end{equation}
 where $T_1=Q\,\sin^2 \beta_1$ and the corresponding distribution of $T_2$ follows from \eqref{B13}
 in result of  $P_{12}$ permutation. It should be perhaps clarified
 that this permutation in general changes the
 kinematic rotation angle that enters the RR coefficients and in consequence the
 $T_2$ distribution does not have to be the same
 as that given in \eqref{B13}.
 \section{Comparison with experiment and conclusions}
 In the preceding section our considerations have been quite general and now
 we wish to apply this theory to the investigation of FSI effects in the
 $pp\to pp\eta$ reaction close to threshold. To simplify matters
 we assume that the excitation energy is sufficiently low so that
 the total orbital momentum in the final state is zero and the
  $\mbox{}^3 P_0\to\mbox{}^1 S_0\, s$ transition amplitude is
  the sole contributor to the cross section. We label the two protons
  as 1 and 2, respectively, and the meson as 3 choosing Jacobi set 3
  for all computations. With the two protons in a singlet state,
  Pauli principle requires that in this frame the orbital
  momenta $\ell_3, \lambda_3$  take only even values. This guarantees that
  the spatial part wave function will be symmetric with respect
  to the $\{12\}$ permutation. We have tried several possible forms
  of the pp potential in the $^1S_0$ state: a delta-shell, a Gaussian
  or a fully realistic Reid potential (for details cf. \cite{deloff}).
  The $\eta$-p interaction is poorly known, for the real part  of the scattering 
  length values between 0.2 and about 1.0 fm have been suggested \cite{GW}.
  Additional difficulty stems from the multi-channel nature of this interaction:
  already at threshold there are open channels so that $\eta$p scattering length
  is a complex number. At present there is not enough information to include these
  additional channels into our formalism, therefore in this work $\eta$-p
  interaction has been simulated by the simplest non-absorptive delta-shell potential
  operative in the s-wave only whose range has been arbitrarily fixed to
  be 1 fm. The depth of this potential can be then adjusted as to yield an assigned
  value of the $\eta$p scattering length $a_{\eta p}$.
  Since the precise value of the latter is not available \cite{GW} we used  
  three values 0.5, 1.0 and 1.5 fm, respectively,
  which we believe is a representative sample. 
  \par
\begin{figure}[!htb]
  \begin{center}
   \includegraphics[scale=0.35]{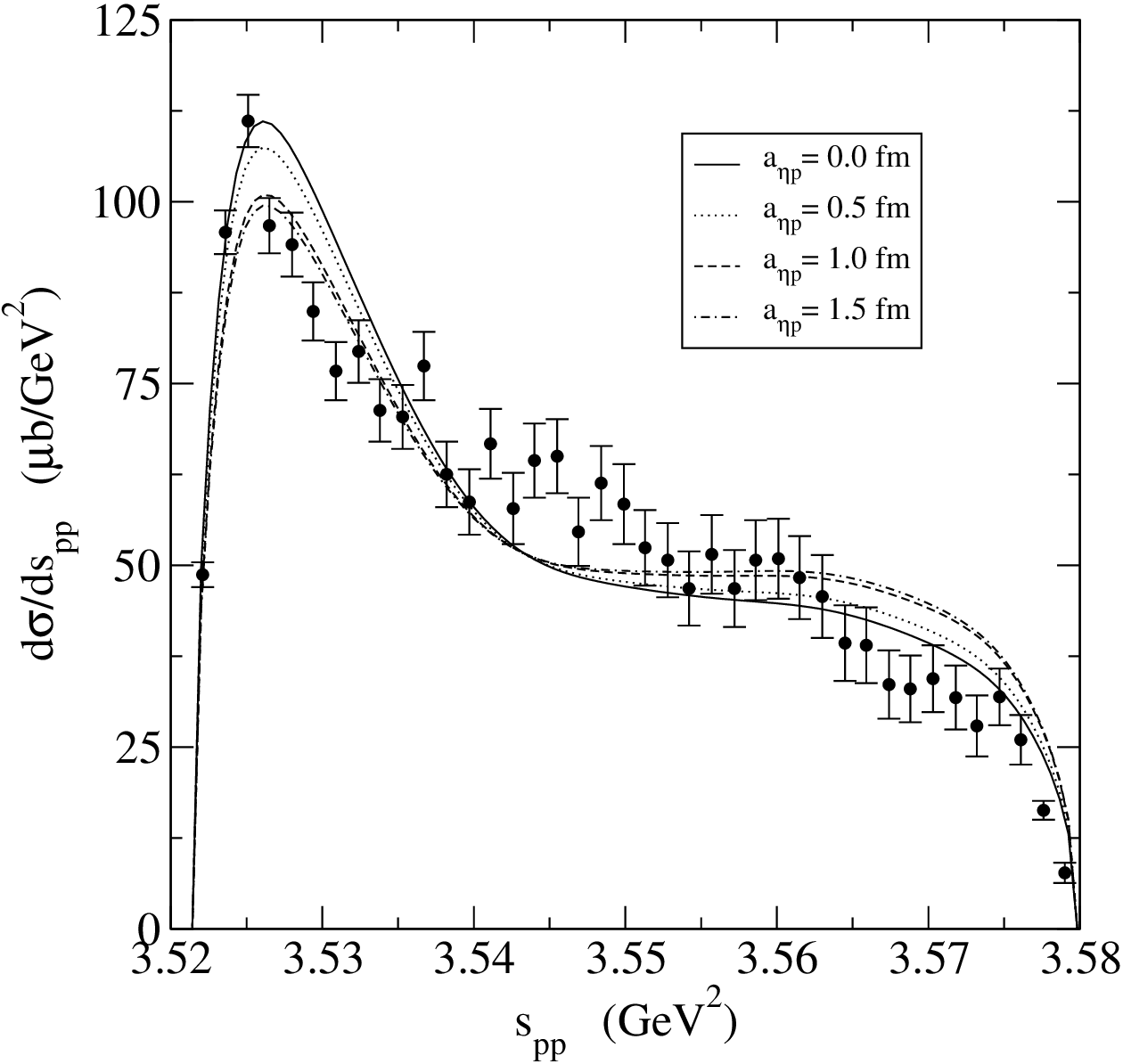}
 \hspace*{5mm}
   \includegraphics[scale=0.35]{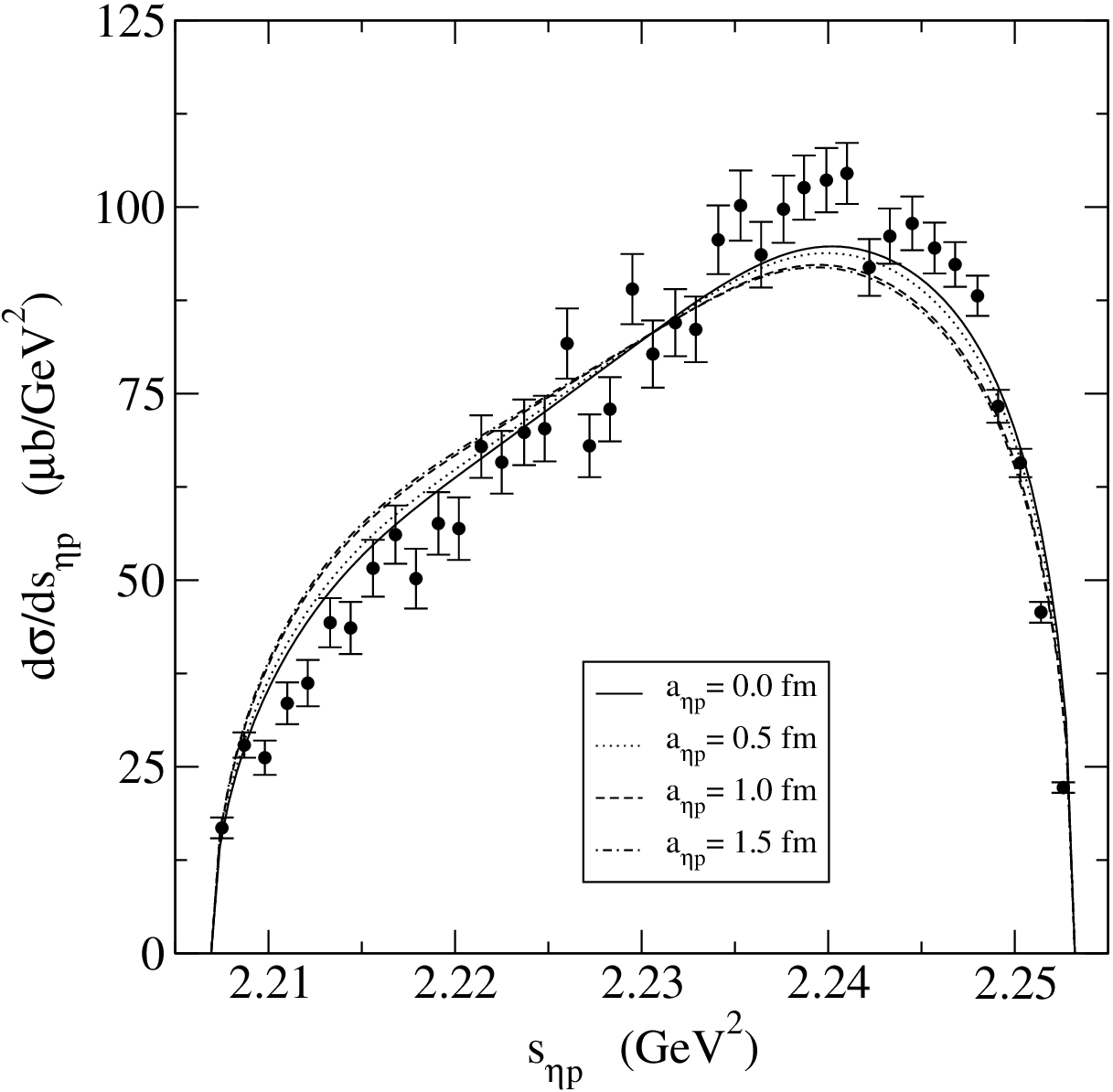}
 \caption{ \small 
 Invariant mass distributions at Q=15.5 MeV
 obtained from a three-body calculation: left panel pp pair;
 right panel $\eta$p pair.
 The data are from \cite{moskal}.}
 \label{three_body} 
 \end{center}
 \end{figure}

  All our computations have been carried out for the lowest 
  excitation energy equal Q=15.5 MeV for which two-particle invariant mass 
  spectra are experimentally available. The system of equations \eqref{A10}
  was perpetually solved by increasing successively in each step the value
  of $K_{max}$ until convergence has been attained. For delta-shell and
  Gaussian pp potentials this occurred at $K_{max}=24$ but with Reid 
  potential this figure must be doubled. Similarly as in the two-body
  case \cite{deloff} the results are completely insensitive to the shape of
  the pp potential.
  The results of our computations are presented in  Fig. \ref{three_body}
  where they are compared with the data from \cite{moskal}.
 Not unexpectedly,
 the dominance of the pp interaction is apparent from both plots.
 Since the values of $a_{\eta p}>1\,fm$ are probably rather unrealistic
 \cite{GW}, we may say that the $\eta$p interaction is of marginal importance
 as far as invariant mass distributions are concerned.
 \par
  Comparing the invariant mass spectra obtained
 in the two-body approach (Fig. \ref{two_body}) with those resulting from the full three-body
 calculation (Fig. \ref{three_body}) we can see that even when
 the $\eta$p interaction is completely disregarded, the invariant mass plots are  different
 in these two cases.
 Although the input in both approaches is the same but the underlying calculational 
 schemes are different. 
 Owing to the proper boundary condition \eqref{B6a} also in absence of $\eta$-p interaction
 the three-body wave function has entangled form, i.e it {\it cannot} be expressed as a 
 product of the p-p wave function times a plane wave associated with the free propagation of the
 $\eta$.
 Since in the approximation using the two-body enhancement factor the very
 existence of the $\eta$  is unaccounted for,  the latter does not  
 depend upon the meson kinematics
 (in \cite{deloff} we made an attempt to lift this deficiency by introducing
 {\it ad hoc} a linear dependence upon q).
 By contrast, even in absence of $\eta$-p
 interaction the three-body enhancement factor accounts for
 the $\eta$ propagation. In particular, 
 the mass of the $\eta$ strongly influences the dynamics of the three-body
 system as can be seen from \eqref{A1:a} and \eqref{A10a}.
 \par
 The  three-body calculation is in both spectra closer to experiment.
 We wish to note here that the curves presented in Fig. \ref{three_body} and
 as well as the dashed curves from Fig. \ref{two_body} 
 contain no adjustable parameters except for the overall
 normalization. Our three-body calculation clearly favors smaller $a_{\eta p}$ values
 $a_{\eta p}\sim\,$ 0\textdiv $0.5\,fm$.  
 For $a_{\eta p}>0.5\,fm$ the height of the close to threshold peak in $s_{pp}$ plot 
 in Fig. \ref{three_body} is depressed which is accompanied
 by a build up of a pronounced shoulder at the high-energy end. 
 At the same time another shoulder appears at the low-energy end in $s_{\eta p}$
 distribution in
 Fig. \ref{three_body}. All these features worsen the agreement with experiment.
 Unfortunately, 
 it would be difficult to improve the existing estimates of
 the $a_{\eta p}$ scattering length
 using the  data displayed in Fig. \ref{three_body},
 especially that absorptive effects have been ignored. 
 \par
 Summarizing, we have developed a general three-body formalism for calculating
 FSI effects in a three-particle final state.
  To the best of our knowledge, this is the first attempt in the literature 
  to derive a three-body final-state interaction enhancement factor.
 The presented calculational framework employs the hypersherical
 harmonics method and is applicable for 
 three-body systems in which no binary bound state can exist.
 The necessary input which must be provided are all three pairwise potentials.
 Detailed computations carried out for the $pp\to pp\eta$ reaction
 at Q=15.5 MeV show that in the invariant mass spectra 
 the role of the $\eta$-p interaction is marginal and more important is
 the proper boundary condition imposed on the final-state wave function.  
 The three-body calculations involving no adjustable parameters  
 reproduce quite well the experimental invariant mass spectra. 
 \par
 Partial support under grant KBN 5B 03B04521 is gratefully acknowledged.
 

\newpage
 \begin{thebibliography}{99}
 \bibitem{review}
       P. Moskal et al.,
       Prog. Part. Nucl. Phys. {\bf 49} (2002) 1.

 \bibitem{reviewt}
           C.~Hanhart, e-Print Archive: hep-ph/0311341.

 \bibitem{zupran}
         M. Abdel-Bary et al., Eur. Phys. J. {\bf A 16} (2003) 127.

 \bibitem{moskal} 
         P.~Moskal et al., Phys. Rev. {\bf C 69} (2004) 025203.
  

 \bibitem{nakayama}
         K. Nakayama et al., Phys. Rev. {\bf C68} (2003) 045201.

 \bibitem{deloff}
         A.~Deloff, Phys. Rev. {\bf C 69} (2004) 035206.

 \bibitem{Fermi}
 E.~Fermi, Nuovo Cim., Suppl. Vol. II, No. 1 (1955) 17\\  
   K.M.~Watson,
   {\it Phys.\ Rev. \ } {\bf 88} (1952) 1163;
   {\it ibid.} {\bf 89} (1953) 575;\\
    A.B.~Migdal, Zh. \ Exper. \ Theor. \ Fiz. \ {\bf 1} (1955) 17.    

 \bibitem{Gold} M.L.~Goldberger and K.M.~Watson, {\it Collision Theory}, Wiley, N.Y.,
         1964;\\ 
 M.~Froissart and R.~Omnes, in {\it Physique des Hautes Energies},\\
 C.~DeWitt and M.~Jacob (eds.), Gordon \& Breach, New York 1965;\\  
 J.~Gillespie, {\it Final state interactions}, Holden-Dey, San Francisco 1964.

 \bibitem{Abram}
 M.~Abramowitz  and I.A.~Stegun (eds.)
 {\it Handbook of Mathematical Functions}, 
 Dover, New York, 1965. 
 
 \bibitem{HH} B.V.~Danilin and M.V. Zhukov, Phys. \ At. \ Nucl. \ {\bf 56} (1993) 460.

 \bibitem{HHH} E. Nielsen, D.V. Fedorov, A.S. Jensen and E. Garrido, Phys. \ Rep. \ 
         {\bf 347}  (2001) 373. 
 
 \bibitem{RR} J.~Raynal and J.~Revai, Nuovo Cim.\ {\bf A 68} (1970) 612.

 \bibitem{GW} A.M.~Green and S.~Wycech, Phys. Rev.  {\bf C 55} (1997) 2167;
         {\it ibid.} {\bf C 60} (1999) 035208.
 \end{thebibliography}
 \end{document}